\documentclass[11pt]{article}
\usepackage{moriondpp,epsfig}

\begin{document}

\vspace*{4cm}
\title{Latest Results from K2K\footnote{Talk presented at the XXXVIth 
 Rencontres de Moriond "Electroweak Interactions and Unified Theories",
 Les Arcs, Savoie, France, March 10-17, 2001.}}


\author{Takanobu Ishii\\(for the K2K Collaboration)}
\address{Institute of Particle and Nuclear Studies, KEK,
  Tsukuba, Japan}

\maketitle\abstracts{
The KEK-to-Kamioka long-baseline neutrino experiment (K2K) 
has begun its investigation of neutrino oscillation, 
and has established the method of a long-baseline neutrino experiment.  
From the first 100 days of data-taking, a deficit of $\nu_{\mu}$ 
in the 1-GeV energy region after 250-km flight was observed 
at the 90\% significance level.}

\section{Introduction}
The possible existence of neutrino masses is the first signal beyond 
the standard model, which suggests grand unification.  
The KEK-to-Kamioka long-baseline neutrino experiment (K2K)~\cite{K2K} 
is motivated by the atmospheric neutrino anomaly found by the Kamiokande
experiment~\cite{Kamiokande}.  
One possible explanation of the anomaly is neutrino oscillation.  
Later, the Super--Kamiokande (SK) group confirmed the deficit of 
upward-going atmospheric $\nu_{\mu}$, and announced evidence for 
the oscillation of atmospheric neutrinos~\cite{SK} with a difference 
of the squared masses, $\Delta m^{2}\sim 3\times 10^{-3}\rm eV^{2}$, and 
the mixing parameter, $\rm sin^{2}2\theta \sim 1$.  
For a neutrino energy of $E_{\nu}$(GeV) and a flight length of $L$(km), 
the oscillation probability is expressed by the following formula 
in a two-flavor approximation: 
\begin{equation}
P(\nu_{\alpha} \rightarrow \nu_{\beta}) = 
\rm sin^{2}2\theta\cdot sin^{2}\frac{1.27\it\Delta m^{2}\cdot L}{\it E_{\nu}}. 
\end{equation}
In order to pin down a small $\Delta m^{2}$, we need a long baseline.  
K2K aims to establish neutrino oscillations in the $\nu_\mu$ disappearance 
mode and in the $\nu_e$ appearance mode, with a well-defined flight length 
and a well-understood flux of a pure $\nu_\mu$ beam.  

\section{Experimental Setup}
K2K uses the SK as a far detector, which is situated 250 km 
from KEK. It is a 50 kton water Cherenkov detector, which is divided into 
an inner part and an outer part.  The inner part is used as 
a target and to measure the neutrino interactions, 
while the outer part is used to veto incoming activities. 
The background for 
the K2K experiment is atmospheric neutrinos of about 8 events/day, 
which can be reduced by a factor of $10^{-6}$ by employing a timing window.  

\begin{figure}[bhtp]
\centerline{\epsfig{file=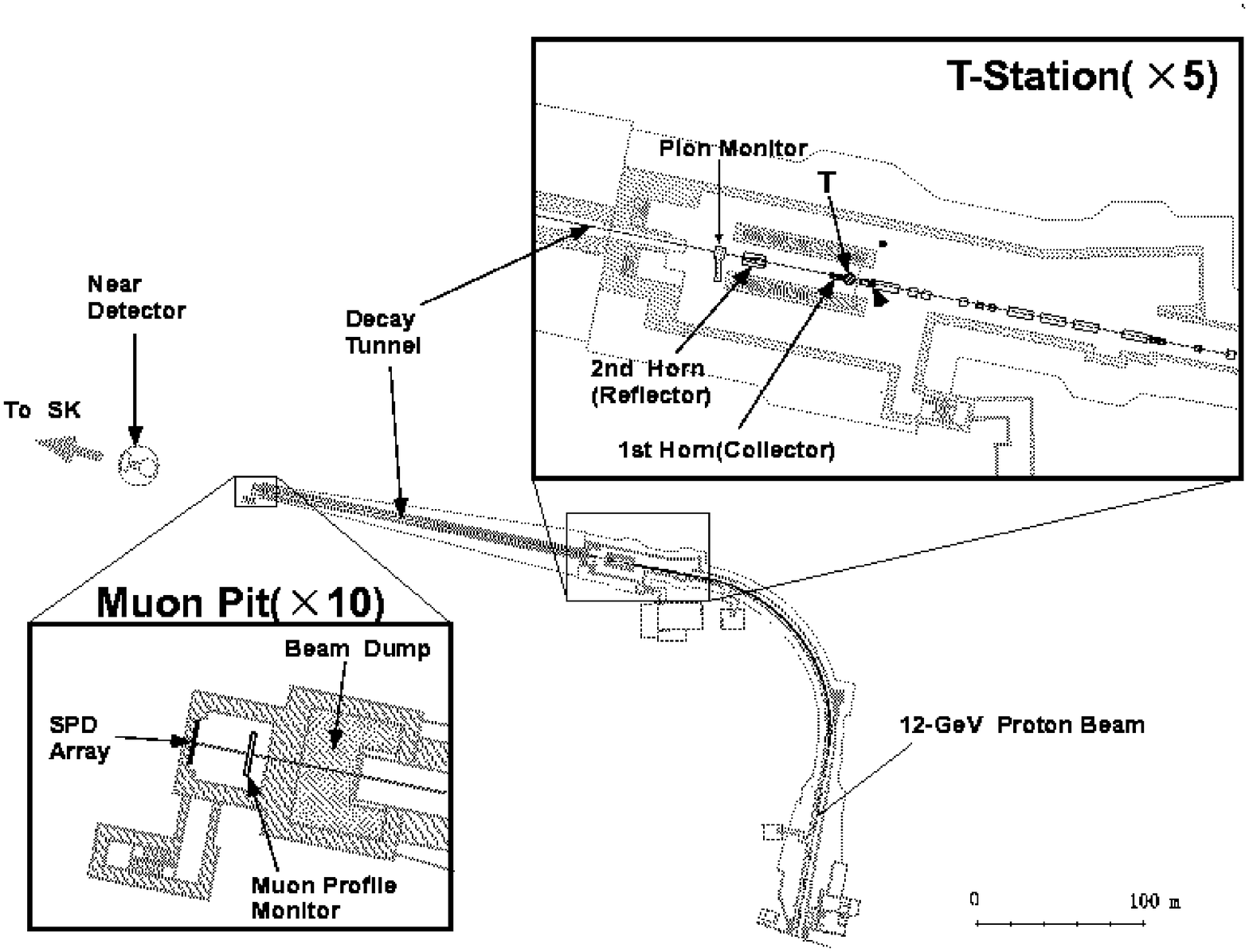,width=11cm}}
\vspace{-0.4cm}
\caption{Near site setup at KEK.}
\label{fig:beamline}
\end{figure}

The setup at the near site is shown in Fig.~\ref{fig:beamline}.  
Every 2.2 sec, a 12-GeV proton beam is fast-extracted from the KEK proton 
synchrotron (PS), making a 1.1-$\mu$sec spill structure.  
The designed intensity is $6\times 10^{12}$ protons/spill. 
Downstream of the arc where the proton beam is bent to the SK direction, 
a pair of horn magnets operating at 250 kA is located.  
The first horn magnet incorporates a pion production target 
comprising a 66-cm-long aluminum cylinder with a 30-mm diameter.  
The horn system focuses positive pions and enhances the neutrino flux 
by a factor of about 20. 
The beam line is aligned toward SK using the global positioning system 
(GPS).  The accuracy of the GPS survey 
is better than 0.01 mrad and the accuracy of the civil construction 
is better than 0.1 mrad.  
For a long-baseline experiment, beam monitors are indispensable.  
We have a pion monitor just after the 2nd horn magnet, a muon monitor 
behind the beam dump, which is located most downstream of 
the 200-m decay tunnel where the pions decay to $\nu_{\mu}$ and muons, 
and a set of near neutrino detectors at 300 m from 
the pion production target.  
\begin{figure}[b!]
\centerline{\epsfig{file=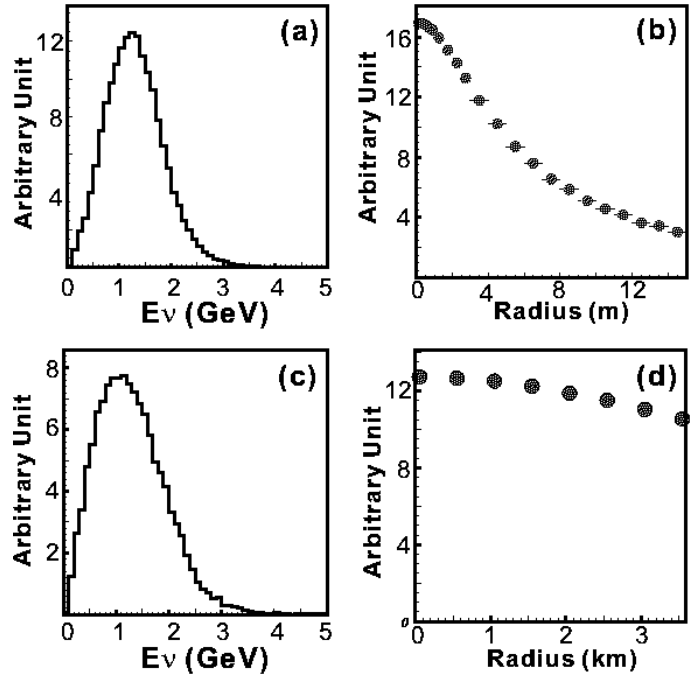,width=9cm}}
\vspace{-0.3cm}
\caption{(a) Expected $\nu_\mu$ spectrum and (b) radial dependence of
the flux at 300 m. 
(c) Expected $\nu_\mu$ spectrum and (d) radial dependence of
the flux at 250 km.}
\label{fig:mcflux}
\end{figure}

The neutrino spectra and the radial distributions at the near site 
and the far site from MC calculations are shown in Fig.~\ref{fig:mcflux}.  
At the far site, the flux is almost constant up to 3 mrad ($\sim 750$ m).  
This is the requirement of beam pointing.  As for the spectra, 
there is a difference between far and near, making a simple 
extrapolation impossible.  
The pion monitor is used to deal with this fact. 

\begin{figure}[bt!]
\centerline{
\epsfig{file=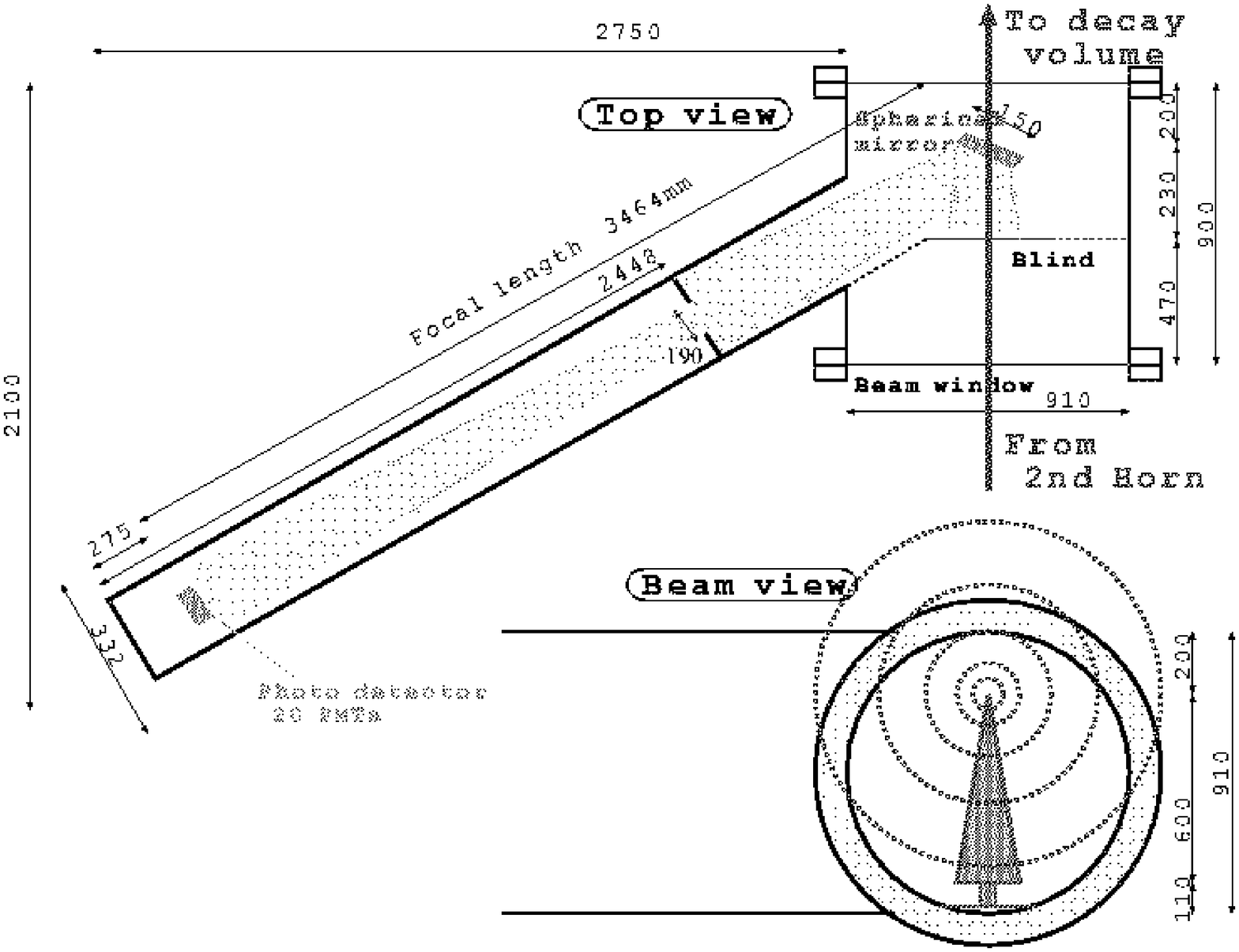,width=8.0cm}
\hspace{1cm}
\epsfig{file=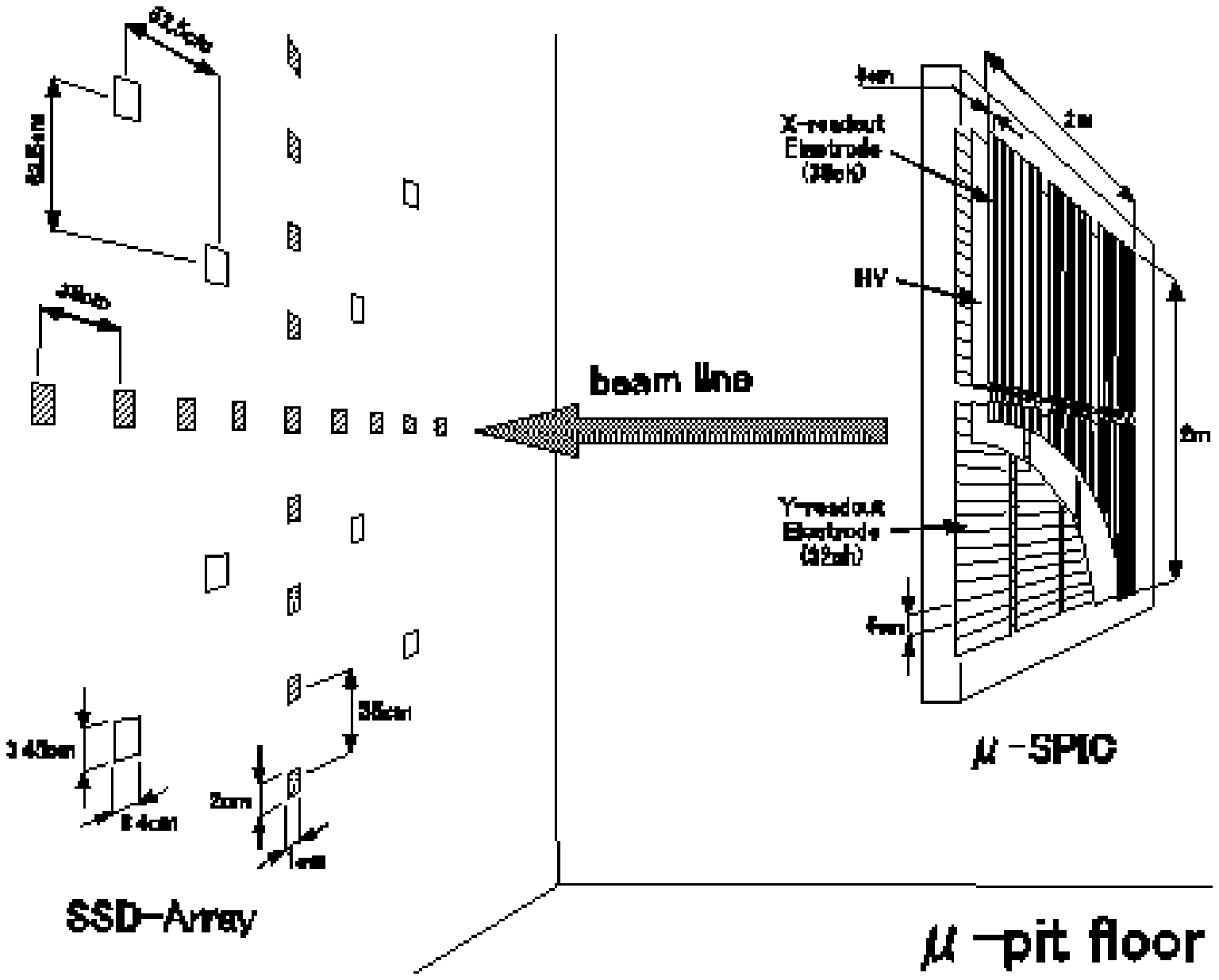,width=7.0cm}
}
\vspace{-0.4cm}
\caption{(Left) Schematic view of the pion monitor. 
(Right) Schematic view of the muon monitor.}
\label{fig:mon}
\end{figure}
The pion monitor~\cite{pimon} measures the momentum and 
angular distributions of pions, 
which are the source of neutrinos.  
Once we know the pion momentum and angular distributions, we can 
calculate the neutrino flux at any distance.  
As a result, we can obtain the far-to-near flux ratio reliably.  
As shown in Fig.~\ref{fig:mon}(Left), the pion monitor is 
a gas Cherenkov detector, which measures the 
Cherenkov ring of pions.  It is occasionally put in 
the beam line.  
A pie-shaped spherical mirror focuses photons onto a PMT array 
located at the focal plane.  
The photon distribution on the PMT array 
is a superposition of slices of the Cherenkov rings from 
pions of various velocities and angles.  
Measurements are made at several indices of refraction.  
The pion two-dimensional distribution of momentum versus angle 
is derived by unfolding the photon distribution data at various 
indices of refraction.  
In order to avoid background from 12-GeV protons 
which have not interacted in the target, the pion monitor is sensitive 
to pion momentum higher than 2 GeV, corresponding to a neutrino energy 
higher than 1 GeV.  

The muon monitor measures the muon profile and intensity.  
Since the muons are the decay partners of the neutrinos from the pions, 
muon measurements give information on the neutrino beam direction 
and intensity.  Since the rate of the muon monitor is high, it can 
measure the beam stability on a spill-by-spill basis.  
The beam direction is tuned by monitoring the muon profile 
at the muon monitor.  
As shown in Fig.~\ref{fig:mon}(Right), 
the muon monitor consists of a segmented ionization chamber and 
a silicon pad detector array.  

\begin{figure}[bt!]
\centerline{
\epsfig{file=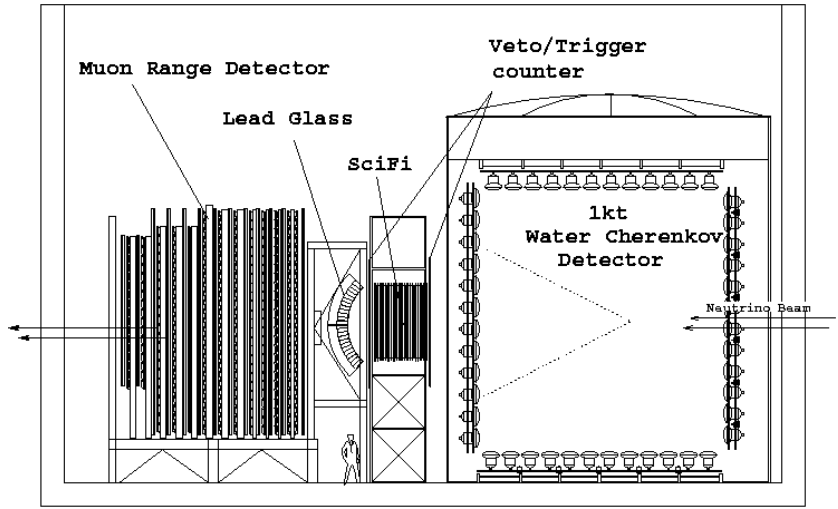,width=11cm}
}
\vspace{-0.4cm}
\caption{Near neutrino detectors.}
\label{fig:k2kfd}
\end{figure}
The near detectors measure the neutrino flux, spectrum and profile, 
studies neutrino interactions at 1-GeV region and measures the $\nu_{e}$ 
contamination.  
As shown in Fig.~\ref{fig:k2kfd}, 
the near detectors consist of a 1kton water Cherenkov detector (1kt), 
a scintillating fiber detector (SCIFI)~\cite{SCIFI}, lead glass counters 
and a muon range detector (MRD)~\cite{MRD}.  
The MRD consists of iron 
plates interleaved with drift tube layers.  

\section{Results}
At this intermediate stage of the K2K experiment, we just compare 
the numbers of events expected and observed at SK.  The expected number 
of events at SK is calculated based on the observed numbers at 
the near detectors, and extrapolated by the far-to-near ratio from MC, 
which is validated by pion monitor measurements.  
The 1kt is used for normalizing the event 
numbers at the near site.  
Since the 1kt is a miniature 
of SK, we can expect the least systematic error in calculating the expected 
number at SK.  
By utilizing iron as 
the target, MRD gives a high event rate, which is suitable for monitoring 
the neutrino beam direction, intensity and spectrum.  
The SCIFI is used to study the neutrino interaction in detail.  

\begin{figure}[!tb]
\centerline{\epsfig{file=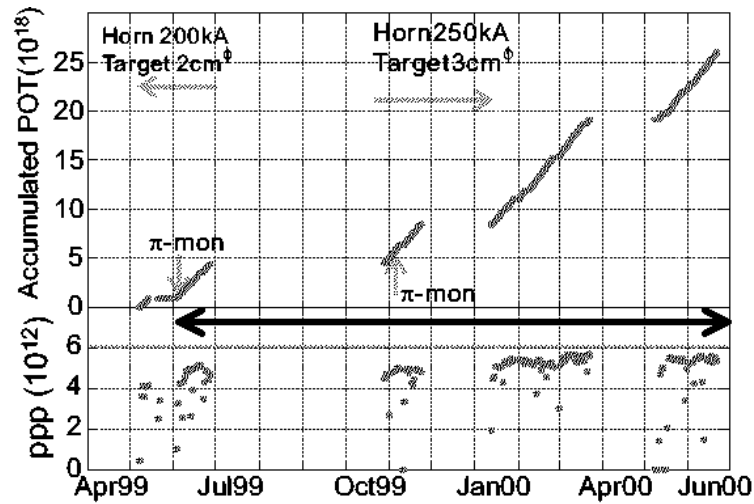,width=10cm}}
\vspace{-0.4cm}
\caption{History of the beam delivery.}
\label{fig:beam99}
\end{figure}

The history of the beam accumulation is shown in Fig.~\ref{fig:beam99}.  
After an engineering run, we started data taking in June, 1999. 
The design intensity has been almost 
achieved.  By June 2000, $2.6\times 10^{19}$ protons have been injected onto 
the target.  Among this number, $2.29\times 10^{19}$ 
have been used for the analysis.  

\begin{figure}[hbt!]
\begin{center}
\epsfig{file=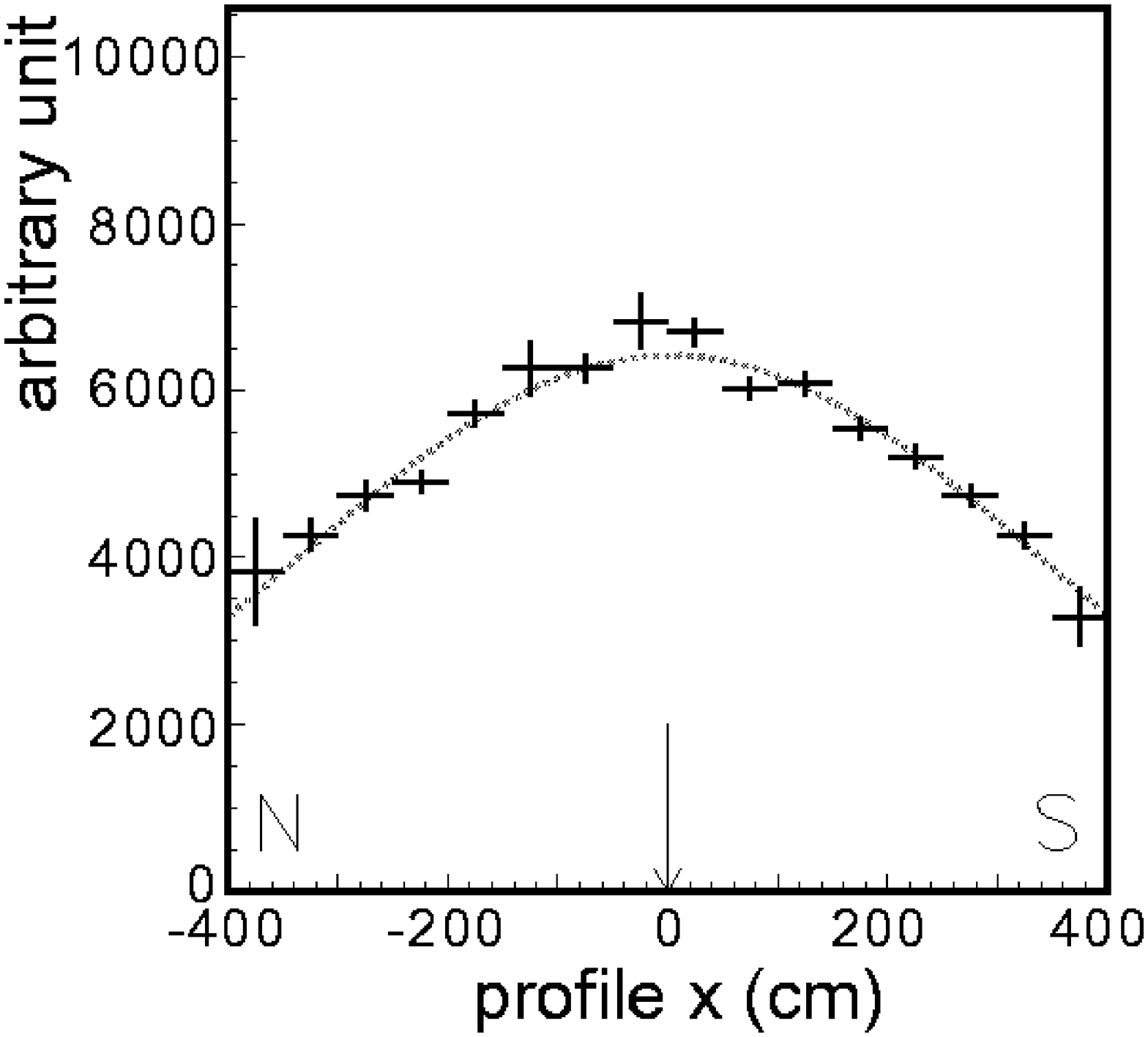,width=7cm}
\hspace{1cm}
\epsfig{file=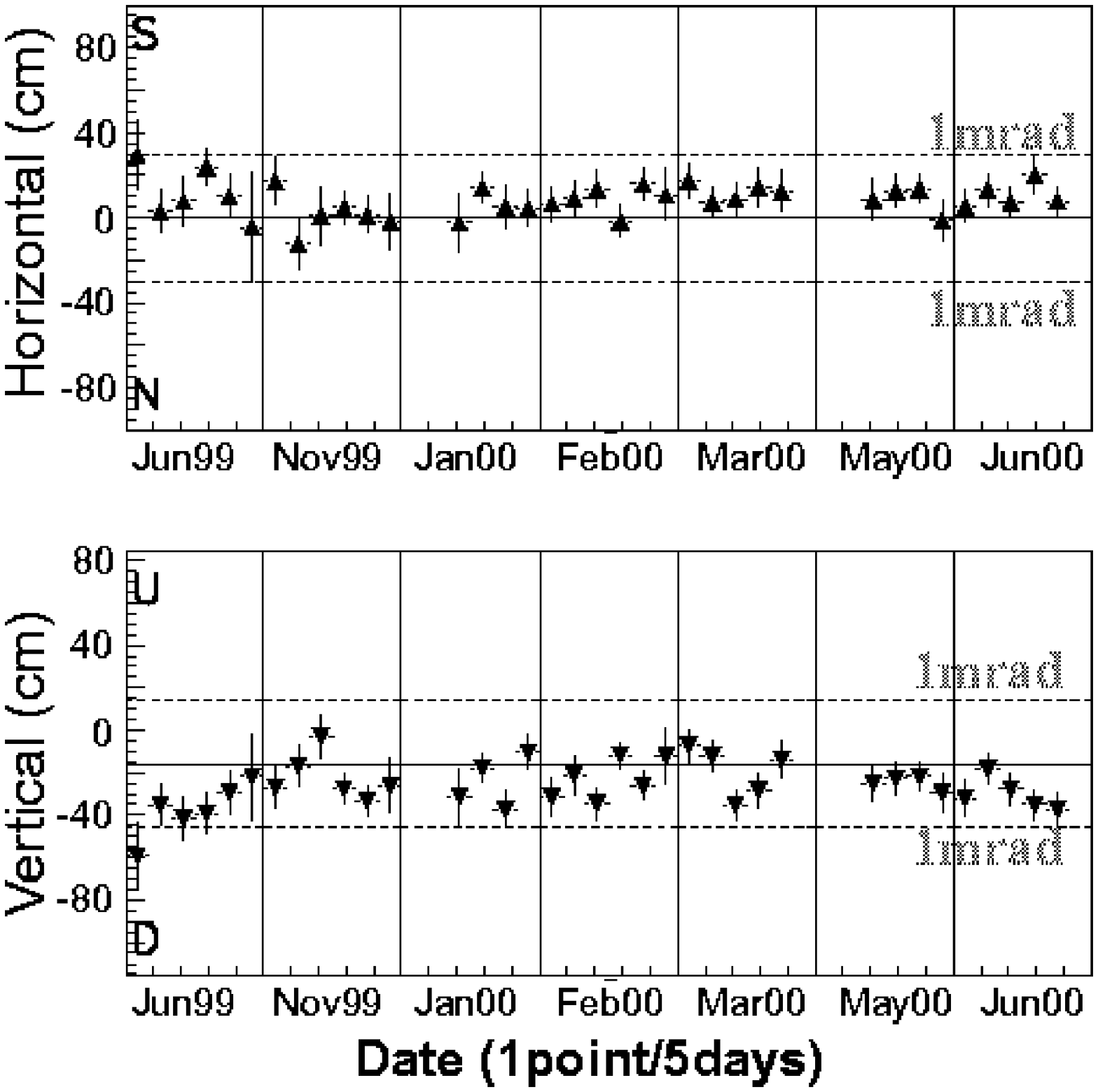,width=7cm}
\end{center}
\vspace{-0.4cm}
\caption{(Left) Beam profile. The points are measured data by MRD 
and the curve is a fitted Gaussian function.
(Right) Stability of the profile center during the experimental periods.}
\label{fig:muplot2}
\end{figure}

The neutrino beam direction is monitored by the MRD. 
Vertex distributions of MRD events in the horizontal and vertical 
directions are shown in Fig.~\ref{fig:muplot2}(Left), 
which gives the neutrino beam profile.  
The center of the profile gives the neutrino beam direction.  
The beam is well directed to SK.  
The center of the profile is plotted in Fig.~\ref{fig:muplot2}(Right) 
as a function of time 
for the horizontal direction and for the vertical direction.  
The beam has been pointed to SK 
within $\pm 1$ mrad during the experimental periods.  

The stability of the beam direction is also monitored by the muon monitor. 
The data show that the beam has been directed to SK within $\pm 1$ mrad 
spill-by-spill during all data-taking periods.  

\begin{figure}[hbt!]
\begin{center}
\epsfig{file=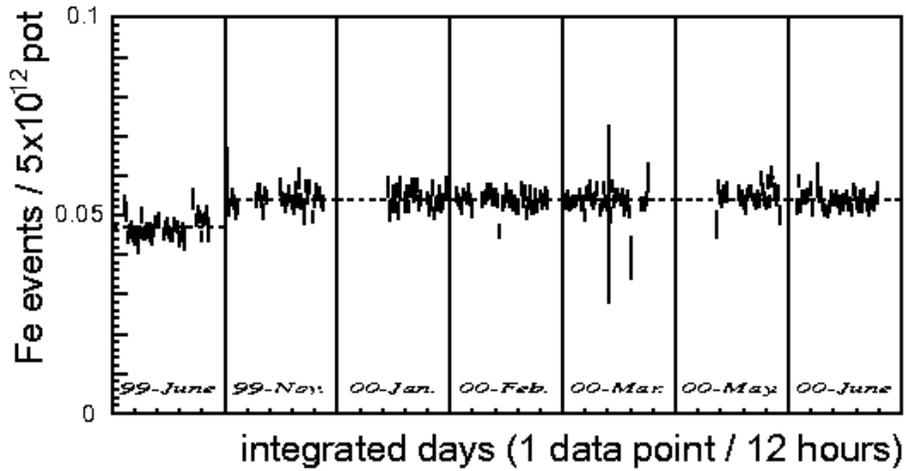,width=12cm}
\end{center}
\vspace{-0.4cm}
\caption{Stability of the event rate measured by MRD.}
\label{fig:stab_rate}
\end{figure}

The stability of the event rate is measured by the MRD 
(Fig.~\ref{fig:stab_rate}).  
The event rate normalized to the proton intensity 
has been stable within the statistical error.  
The slight difference in June, 1999, 
is due to a different horn current and 
a different target size.  

\begin{figure}[hbt!]
\begin{center}
\epsfig{file=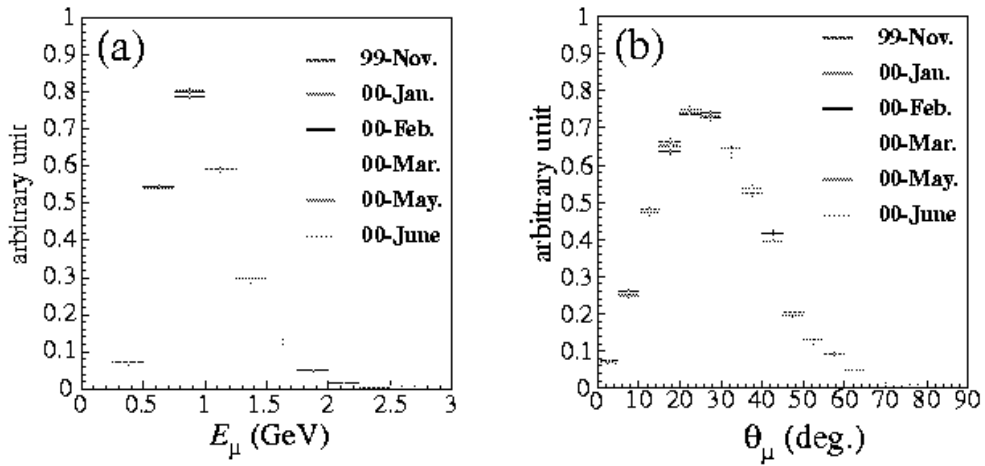,width=13cm}
\end{center}
\vspace{-0.6cm}
\caption{Muon (a) energy spectra and (b)
angular distributions of $\nu_\mu$ interactions in MRD iron
plates for each month.}
\label{fig:stab_emuang}
\end{figure}

The muon energy and angular distributions are also continuously 
monitored.  They are shown in Fig.~\ref{fig:stab_emuang} 
for each one-month period.  
These figures show no change in the energy and angular distributions, 
implying that the neutrino energy spectrum has been stable throughout 
these periods.  

\begin{figure}[bt!]
\centerline{
\epsfig{file=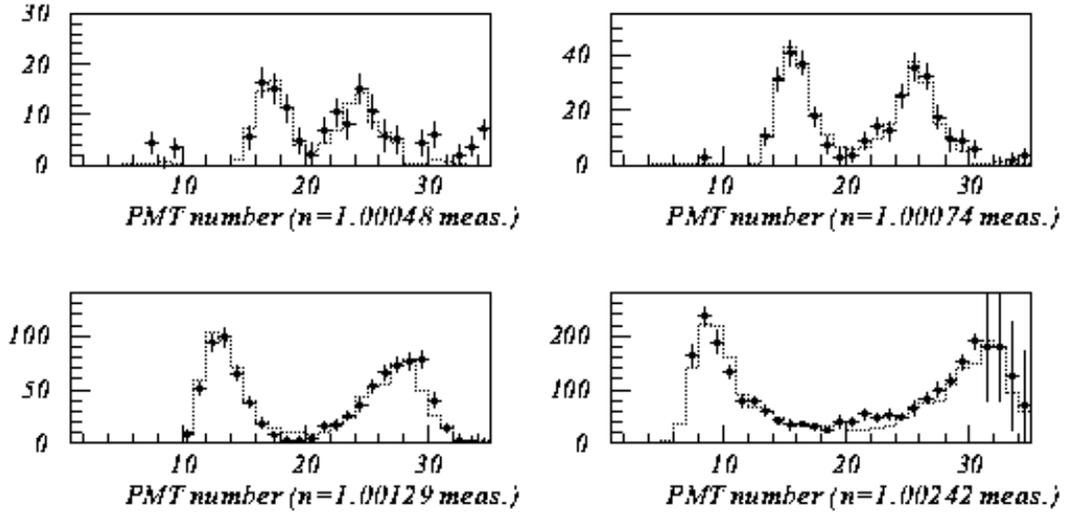,width=14cm}
}
\vspace{-0.4cm}
\caption{Cherenkov photon distributions measured by the pion monitor. 
 The points with error bars show data and the histograms show the 
 fit results.}
\label{fig:pimonfit}
\end{figure}

\begin{figure}[hbt!]
\centerline{\epsfig{file=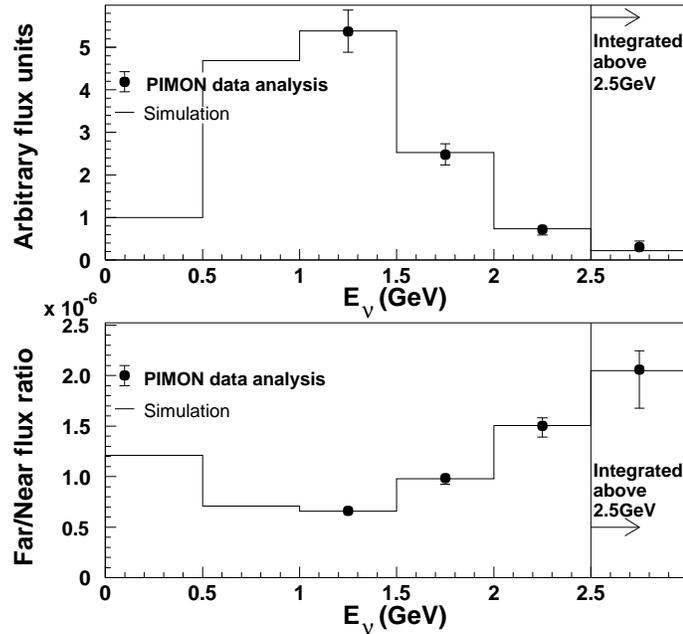,width=9cm}}
\vspace{-0.5cm}
\caption{(Top) $\nu_\mu$ energy spectrum at
the near site (Bottom) Far-to-near
$\nu_\mu$ flux ratio. The histograms are from the beam
simulation results. The data points are derived from the
pion monitor measurement.}
\label{fig:pimon_rat}
\end{figure}

The pion monitor measurements were performed at the beginnings of 
June and November, 1999. 
The measured photon distributions on the PMT array 
are shown in Fig.~\ref{fig:pimonfit} with the fitted results 
for some of indices of refraction. 
Fig.~\ref{fig:pimon_rat}(Top) shows the neutrino energy spectral 
shape at the near site inferred from the pion monitor measurements 
along with the beam simulation result.  
The beam simulation is validated by the pion monitor measurement very well. 
Fig.~\ref{fig:pimon_rat}(Bottom) shows the far-to-near flux ratio.  
The data points are derived 
from the pion monitor measurement and the lines are the beam 
simulation results.  They show very good agreement.  

Based on these stability measurements and the pion monitor measurement, 
the expected number of events at SK can be calculated reliably.  
We obtained $37.8 \pm 0.2(\rm stat) ^{+3.5}_{-3.8}(\rm sys)$ 
based on the 1kt normalization.  
The sources of the systematic error are the uncertainty of 
the far-to-near ratio of 6-7\% coming from the pion monitor measurement, 
the uncertainty due to the 1kt measurement of 5\% coming mainly 
from the fiducial volume error and the uncertainty due to 
the SK measurement of 3\%, also coming mainly from the fiducial 
volume error.  
Calculations based on the MRD events and the SCIFI events give 
$41.0 ^{+6.0}_{-6.6}$ and $37.2 ^{+4.6}_{-5.0}$, respectively, 
which are consistent with the number obtained from 1kt.  

\begin{figure}[hbt!]
\centerline{
\epsfig{file=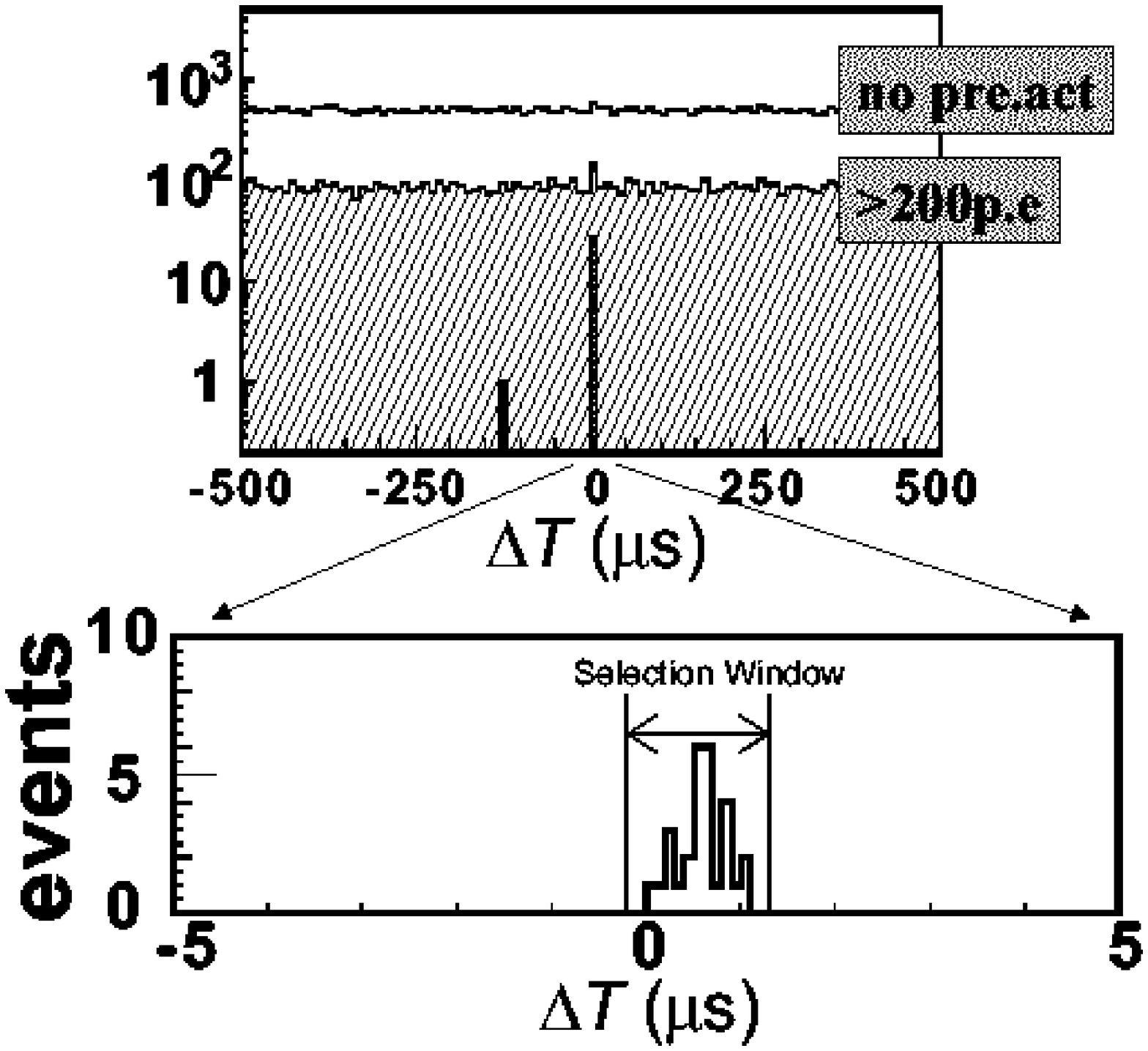,width=7.5cm}
\hspace{1cm}
\epsfig{file=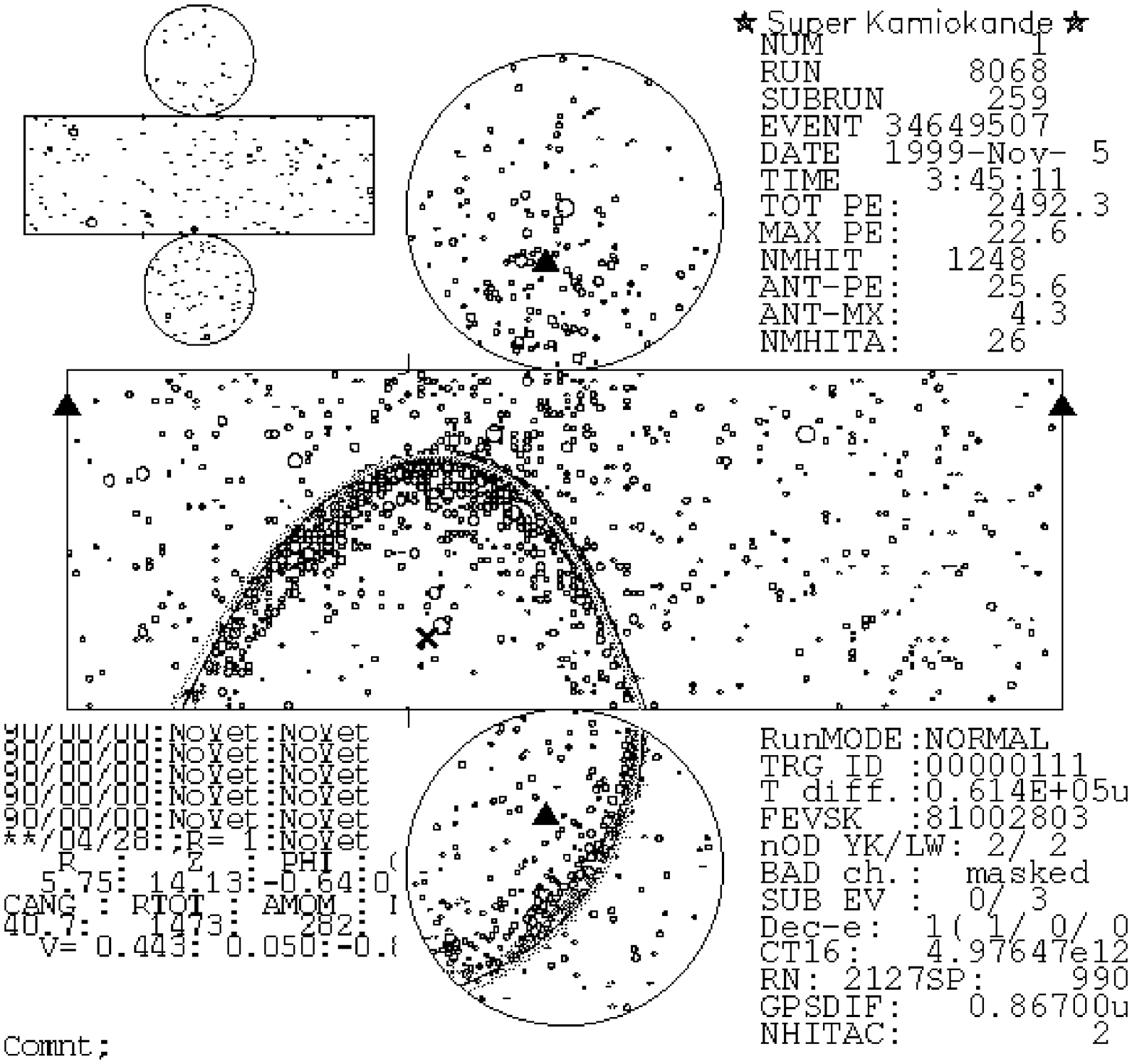,width=7.5cm}
}
\vspace{-0.4cm}
\caption{(Left) $\Delta T$ distribution. 
 (Right) Candidate of FC single ring $\mu$-like event
 in the fiducial volume.}
\label{fig:sktdif-skevt}
\end{figure}

For the SK event analysis, data reduction similar to that used in 
atmospheric neutrino analyses is applied.  
The criteria are: 1) There is no detector activity within 30 $\mu$sec 
before the event.  2) The total collected photo-electrons in 
a 300 nsec time window is larger than 200. 3) The number of PMTs in 
the largest hit cluster in the outer-detector is less than 10.  
4) The deposited energy is larger than 30 MeV.  5) The reconstructed 
vertex is at least 2 m inside the wall, which defines the fiducial 
volume of 22.5 kt.  
The overall detection efficiency of SK is 79\% including neutral 
current interactions.  
The inefficiency is mainly due to the energy cut.  
Accelerator-produced neutrino events at SK are selected by using 
GPS time information.  The distribution of the time difference 
between the SK trigger time and the KEK-PS beam spill start time 
at various reduction stages 
is shown in Fig.~\ref{fig:sktdif-skevt}(Left).  
After all of the event selection cuts, the $\Delta t$ 
distribution is consistent with the beam spill time of 1.1 $\mu$sec 
and GPS resolution of about 100 nsec.  
A typical SK event caused by an accelerator-produced neutrino 
is shown in Fig.~\ref{fig:sktdif-skevt}(Right).  
We observed 28 fully contained (FC) events in the SK 
fiducial volume.  The expected background from atmospheric neutrinos is 
less than $10^{-3}$ events.  
The observed numbers of events at SK are compared with the expected 
numbers in each event category in Table ~\ref{tbl:skexpect}.  
\begin{table}[!hbt]
\caption{Breakdown of the observed and expected numbers of events
into event categories.}
\label{tbl:skexpect}
\vspace{0.2cm}
\begin{center}
\begin{tabular}{|lll|r|c|}
\hline
\multicolumn{3}{|l|}{Category}            & Obs. & Expected     \\
\hline
 & \multicolumn{2}{l|}{1-ring $\mu$-like} & 14 & 20.9           \\
 & \multicolumn{2}{l|}{1-ring e-like}     &  1 &  2.0           \\
 & \multicolumn{2}{l|}{multi rings}       & 13 & 14.9           \\
\hline
\multicolumn{3}{|l|}{Total}               & 28 & 37.8$^{+3.5}_{-3.8}$ \\
\hline
\end{tabular}
\end{center}
\end{table}

\begin{figure}[bt!]
\centerline{
\epsfig{file=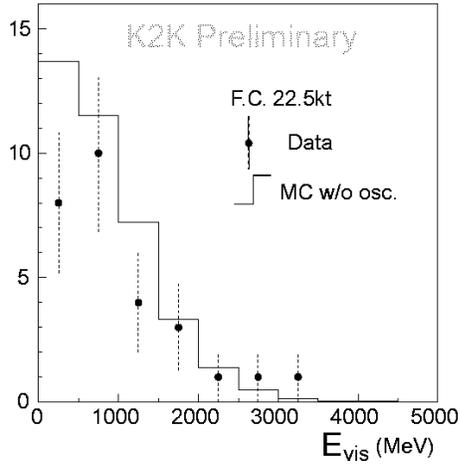,width=6cm}
}
\vspace{-0.4cm}
\caption{Visible energy distribution of the selected 
FC events in the SK fiducial volume.}
\label{fig:visible_e}
\end{figure}

\begin{figure}[bt!]
\centerline{
\epsfig{file=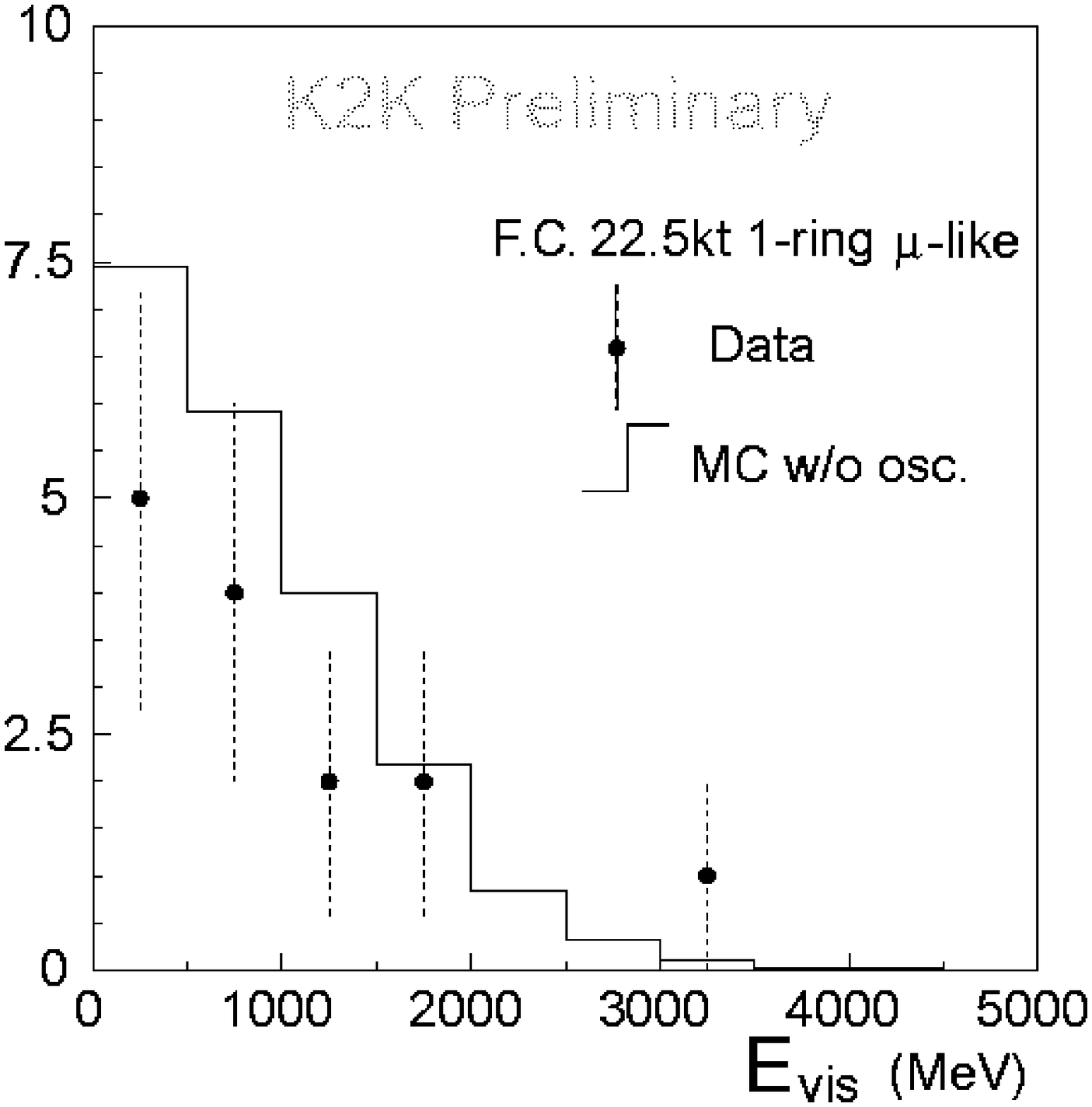,width=6cm}
\hspace{1cm}
\epsfig{file=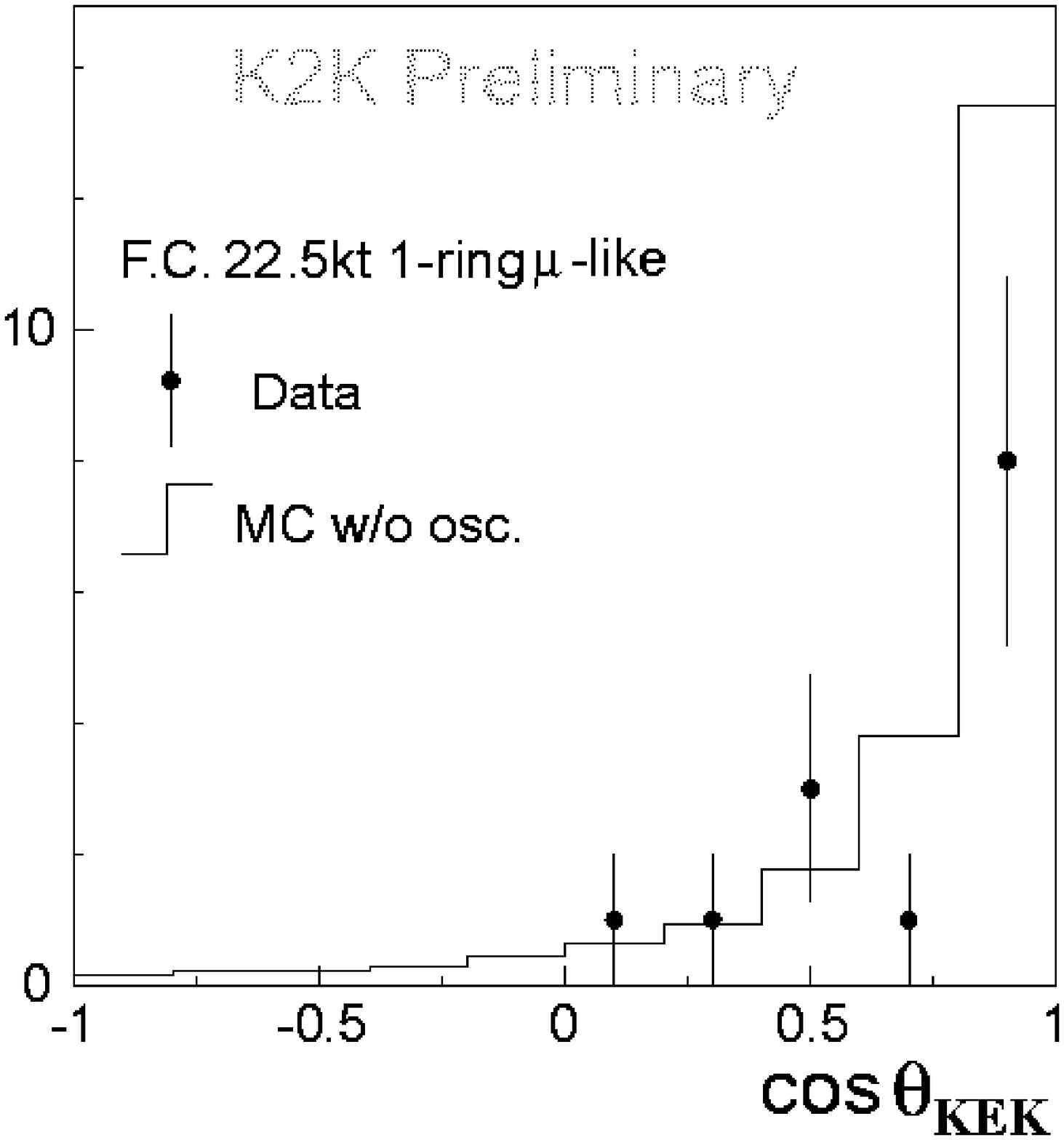,width=6cm}
}
\vspace{-0.4cm}
\caption{(Left) Visible energy distribution of the 1-ring $\mu$-like 
FC events in the SK fiducial volume. 
(Right) Angular distribution with respect to the KEK direction 
of the 1-ring $\mu$-like FC 
events in the SK fiducial volume.}
\label{fig:1ring_mu}
\end{figure}

The visible energy (electron equivalent energy) distribution of 
the selected 28 events is shown 
in Fig.~\ref{fig:visible_e} together with the MC expectation.  
The systematic error of the MC expectation is under study.  
The visible energy distribution and the 
angular distribution with respect to the KEK direction of 
the 1-ring $\mu$-like events are shown in Fig.~\ref{fig:1ring_mu}. 

\section{Conclusions}
The method of a long-baseline neutrino experiment 
has been established, namely beam steering, monitoring the neutrino 
beam at the near site, predicting neutrino properties at 
the far site from the near site measurements and time synchronization 
between the near site and the far site.  
In the data taken from June 1999 to June 2000, 
corresponding to $2.29\times 10^{19}$protons on target, 
we observed 28 FC events in the fiducial volume of SK, 
while the expected number without oscillation was 37.8 with an error 
of about 10\%.  
This means that the deficit of $\nu_{\mu}$ in the 1-GeV energy region 
after 250-km 
flight was observed at the 90\% significance.  

We aim to accumulate data corresponding to at least 
$10^{20}$protons on the target and to perform 
a spectral analysis to see the characteristic energy distortion in 
the case of neutrino oscillation.  We will also study 
the $\nu_{e}$ appearance. 
The experiment resumed in January 2001 and we are taking 
data until July for this year.

\end{document}